# Algorithm 985: Simple, efficient, and relatively accurate approximation for the evaluation of the Faddeyeva function

MOFREH R. ZAGHLOUL, United Arab Emirates University[1]

We present a new simple algorithm for efficient, and relatively accurate computation of the Faddeyeva function $w(z)$. The algorithm carefully exploits previous approximations by Hui et al [1978] and Humlíček [1982] along with asymptotic expressions from Laplace continued fractions. Over a wide and fine grid of the complex argument, $z=x+iy$, numerical results from the present approximation show a maximum relative error less than $4.0 \times 10^{-5}$ for both real and imaginary parts of $w$ while running in a relatively shorter execution time than other competitive techniques. In addition to the calculation of the Faddeyeva function, $w$, partial derivatives of the real and imaginary parts of the function can easily be calculated and returned as optional output.

---

Author's addresses: M. Zaghloul, Department of Physics, College of Sciences, United Arab Emirates University, Al-Ain, 15551, UAE.

1.  **INTRODUCTION**

The complex probability function, commonly known as the Faddeyeva function, has gained wide interest in the literature due to its importance and application in many fields of physics such as plasma spectroscopy, atmospheric radiative transfer, nuclear physics, nuclear magnetic resonance, etc. A fairly extensive body of the literature has been developed discussing the properties and numerical evaluation of this function, some of them use different names [Faddeyeva and Terent'ev 1961, Young 1965; Armstrong 1967; Gautschi 1969; 1970; Hui et al. 1978; Humlíček 1982; Dominguez et al. 1987; Poppe and Wijers. 1990a,b; Lether and Wenston 1991; Schreier 1992; Shippony and Read 1993; Weideman 1994; Wells 1999; Luque et al. 2005; Letchworth and Benner 2007; Zaghloul 2007; Abrarov et al. 2010a,b; Zaghloul and Ali 2011, Boyer and Lynas-Gray 2014, Zaghloul 2015]. In some practical applications, for example, in the line-by-line calculation of microwave and infrared radiative transfer from spectroscopic data, evaluation of the Faddeyeva function may account for a significant fraction of the execution time and the development of more efficient approximations continues to be a priority.

The function is defined mathematically as the scaled complementary error function for a complex variable and can be expressed as,

$$w(z) = e^{(-iz)^2} erfc(-iz)$$
$$= \begin{cases} \dfrac{i}{\pi} \int_{-\infty}^{+\infty} \dfrac{e^{-t^2} dt}{z-t} & \Im m\ z > 0 \\ \dfrac{i}{\pi} \int_{-\infty}^{+\infty} \dfrac{e^{-t^2} dt}{z-t} + 2e^{-z^2} & \Im m\ z < 0 \end{cases} \quad (1)$$

where $z=x+iy$ is a complex argument and $erfc(z)$ is the complementary error function. The real and imaginary parts of the function are known as the real and imaginary *Voigt functions* $V(x,y)$ and $L(x,y)$, that is,

$$w(x+iy) = V(x,y) + i L(x,y). \quad (2)$$

The real part of the function, $V(x,y)$, is widely used to represent spectral line shapes in many fields of physics such as astrophysics, atmospheric spectroscopy and radiative transfer, plasma physics, etc. The quantity $x$ is the non-dimensional frequency and $y$ is the damping parameter, both expressed in Doppler width units. The imaginary Voigt function $L(x,y)$ is also useful as it allows for the calculation of the spectral line profile including line mixing.

A closed form expression does not exist for the integral defining the Faddeyeva function and as a result a wide variety of algorithms have been developed to evaluate the function numerically. The accuracy and computational speed of these algorithms vary significantly and the choice of the appropriate algorithm will depend on the particular application under consideration. For applications requiring huge numbers of evaluations (sometimes in excess of $10^{11}$) computational speed is the deciding factor and, when low accuracy evaluations are sufficient, implementations of the methods of Hui et al. [1978] and Humlíček [1982] are commonly used. Humlíček's routine was introduced to overcome the inevitable failure of any rational approximation (similar to Hui's algorithm) near the real axis; however, the two routines suffer from a loss of their claimed accuracies near the real axis as shown by many authors in the literature (see, for example, Wells [1999]; Zaghloul and Ali [2011]). Assessments, using Algorithm 916, Zaghloul and Ali [2011] as a reference, indicate that Humlíček's w4 algorithm satisfies its claimed accuracy of $9.6 \times 10^{-5}$ only for $y > 10^{-6}$.

It is also worth mentioning that Hui's algorithm and Humlíček's w4 algorithm form the basis of several other refinements [Schreier 1992; Kuntz 1997; Wells 1999; Imai et al 2010]. Following Karp's [1978] suggestion, Schreier [1992] replaced Hui's rational approximation for the real part of the Faddeyeva function by $V(x,y) = exp(-x^2) + y/(x^2\sqrt{\pi})$ for small $y$ and $y/x^2 < 10^{-4}$. Nevertheless, large errors for intermediate $x$ and small $y$ values were not removed. A new implementation of Humlíček's algorithm for approximating $V(x,y)$, proposed by Kuntz [1997], claims to have improved efficiency over other implementations but the method suffers from a loss of accuracy for small $y$. A number of errors, in the Kuntz' implementation, have been reported in [Ruyten 2004]. Wells [1999] and Imai et al [2010] reintroduced Humlíček's w4 algorithm with redefined boundaries and added a new region (called Region 0) for $|z| \rightarrow \infty$ (i.e., far from the line center) where the simple expression $w(z) \sim i/(\pi z)^{1/2}$ gives adequate relative accuracy. Imai's algorithm for $V(x,y)$ also loses accuracy for small $y$ while Wells' method replaces the w4 rational approximation in Region IV by the less efficient CPF12 algorithm [Humlíček 1979] to overcome this problem. Further, Wells' code optionally provides the imaginary part of the Faddeyeva function and its derivatives. As well as being less efficient, the claimed accuracy $10^{-5}$ of Wells' algorithm is not satisfied in some regions of the computational domain. In most of the above studies, Humlíček's w4 algorithm is shown to be remarkably efficient when the parameter $x$ is a vector and $y$ is a scalar. A brief survey and benchmarking tests by Schreier [2011] reported that, for *Fortran* and *Python* implementations, programming language, compiler choice, and implementation details influence computational speed to such a degree that there is no unique ranking of algorithms.

In this paper we present a simple and relatively accurate approximation of the Faddeyeva function that overcomes the loss of accuracy problem mentioned above, while running in a competitive execution time. The new approximation is particularly suitable where the evaluation of the Faddeyeva function may account for a significant part of the execution time. In Section 2 we provide details about the proposed routine and its accuracy while implementations in *Matlab* and *Fortran* together with performance results are presented and discussed in Section 3.

## 2. THE ROUTINE AND ITS ACCURACY

The Faddeyeva function can be approximated asymptotically by the Laplace continued fractions [Faddeyeva and Terent'ev 1961, Abramowitz and Stegun 1964, Gautschi 1970],

$$w(z) \cong \frac{i}{\sqrt{\pi}} \, \frac{1}{z-} \, \frac{1/2}{z-} \, \frac{1}{z-} \, \frac{3/2}{z-} \, \frac{2}{z-} \ldots \ldots \ldots \quad (3)$$

where the continued fraction needs to be truncated at some convergent for practical evaluation. For large values of $|z|$ a few convergents may be used to obtain the desired accuracy.

Using Algorithm 916 as a reference, the regions of applicability vs the number of convergents retained (up to four convergents) are specified for an accuracy better than $4.0 \times 10^{-5}$. However, as the computational cost (execution time) increases with the number of convergents retained to approximate the function, the minimum number of convergents necessary to approximate the function to within the specified accuracy is chosen wherever these regions overlap. The first four rows in Table 1 summarize the proposed regions of applicability and the equivalent rational approximation up to four convergents of Laplace continued fractions. As can be seen from the table, these convergents can be used to approximate the Faddeyeva function to accuracy better than $4.0 \times 10^{-5}$ for $|z|^2 \geq 30$ except for the narrow strip defined by $62.0 > |z|^2 \geq 30.0$ & $y^2 < 10^{-13}$.

For the region $|z|^2 < 30.0$ and except for the strip given by the intersection of $30.0 > |z|^2 > 2.5$ and $y^2 < 0.072$, tests using a fine grid showed Hui's p6 approximation (rational approximation with the degree 6 polynomial in the numerator and degree 7 polynomial in the denominator) for both $V(x,y)$ and $L(x,y)$, to approximate the function to accuracy better than the targeted accuracy ($<4.0 \times 10^{-5}$). Finally, the approximation proposed by Humlíček [1982] to be used in Region IV in the w4 algorithm is found to be very accurate for very small values of $y$ and values of $x$ in the region $|z|^2 < 62.0$ [Zaghloul 2015]. The expression was found to give the required accuracy for values of $y^2 < 0.072$. We propose using this expression for the region of small $y$, where the first four convergents of Laplace continued fractions and Hui's approximation cannot secure the targeted accuracy in the region $|z|^2 < 62.0$.

Based on the above findings we propose a simple algorithm for the calculation of the Faddeyeva function with accuracy $<4.0 \times 10^{-5}$. The algorithm employs a partitioning of the computational domain with different simple approximate methods in different regions as discussed above and explained in Figure 1 and Table 1 below.

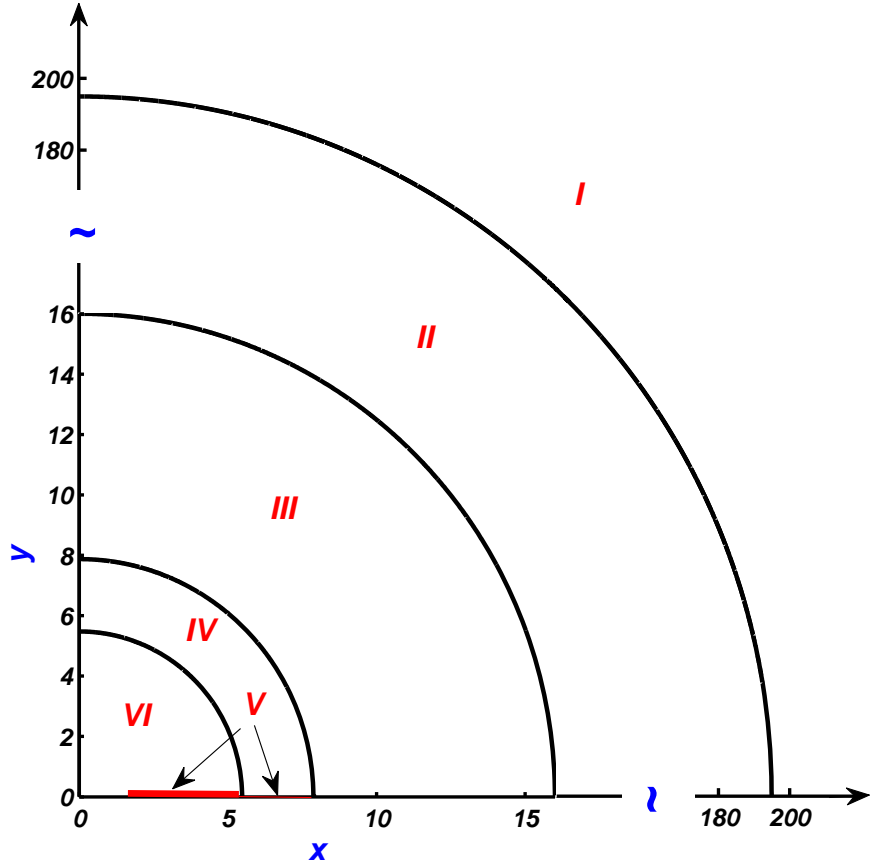

**Figure 1:** Regions of the partitioning of the *x-y* plane in the present proposed routine

**Table 1:** Regions of the proposed partitioning and the corresponding computational method used

| Region | Borders | Method |
|---|---|---|
| I | $\|z\|^2 \geq 3.8 \times 10^4$ | 1 *convergent (continued fractions)* <br> $w(z) \approx i/z\sqrt{\pi}$ |
| II | $3.8 \times 10^4 > \|z\|^2 \geq 256.0$ | 2 *convergents (continued fractions)* * <br> $w(z) \approx iz/\sqrt{\pi}\,(z^2 - 0.5)$ |
| III | $256.0 > \|z\|^2 \geq 62.0$ | 3 *convergents (continued fractions)* <br> $w(z) \approx i(z^2 - 1)/z\sqrt{\pi}\,(z^2 - 1.5)$ |
| IV | $62.0 > \|z\|^2 \geq 30.0\ \&\ y^2 \geq 10^{-13}$ | 4 *convergents (continued fractions)* ** <br> $w(z) \approx iz(z^2 - 2.5)/\sqrt{\pi}\,(z^2(z^2 - 3) + 0.75)$ |
| V | $62.0 > \|z\|^2 \geq 30.0\ \&\ y^2 < 10^{-13}\ \cup$ <br> $30.0 > \|z\|^2 \geq 2.5\ \&\ y^2 < 0.072$ | *Rational approximation for region IV in Humlíček's w4 algorithm* |
| VI | *Otherwise* | *Hui's p6 approximation* |

*, ** The rational approximations resulting from the second and fourth convergents are identical to the approximations for regions I and II in Humlíček's *w*4 algorithm, respectively.

Figure 2 shows surface plots of the absolute relative error, in the calculation of the real, $V$, and imaginary, $L$, parts ($\delta_V = |(V - V_{ref})|/V_{ref}$ and $\delta_L = |(L - L_{ref})|/L_{ref}$), of the Faddeyeva function, resulting from using the present routine with Algorithm 916 as a reference. For the whole domain of calculation, the maximum relative error for both the real and imaginary parts was found to be less than $4.0 \times 10^{-5}$.

Calculating the partial derivatives of the real and imaginary parts of the Faddeyeva function is important for many applications; see [Schreier 1992, Wells 1999, Letchworth and Benner 2007]. Having computed the Faddeyeva function, the partial derivatives may be computed with relative simplicity using the relation

$$w'(z) = \frac{2i}{\sqrt{\pi}} - 2z\,w(z) \tag{4}$$

The partial derivatives of the real part of the function can be calculated and returned as optional output with the Faddeyeva function while the derivatives of the imaginary part of $w$ may be obtained directly from the derivatives of the real part using Cauchy-Riemann relations.

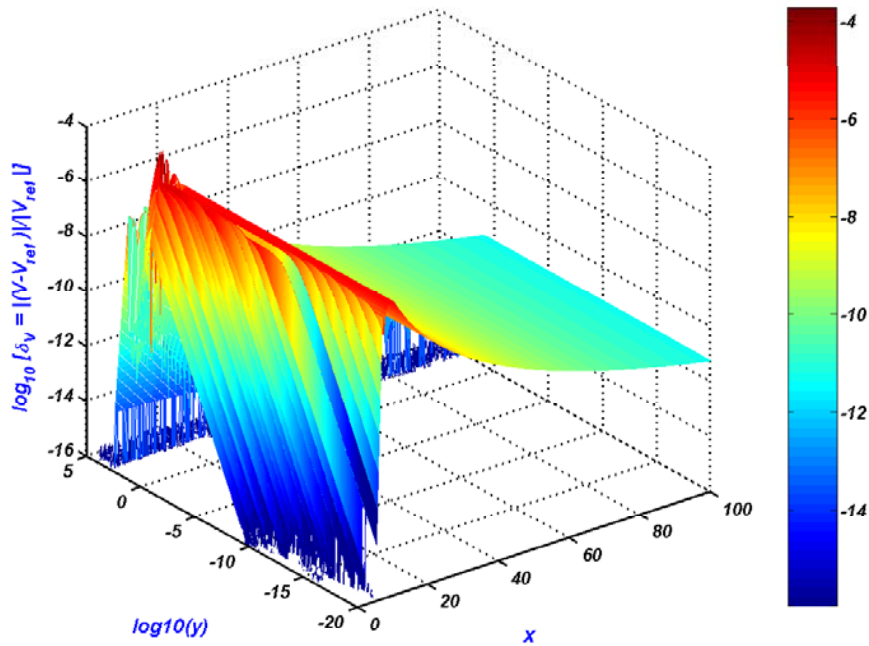

(a)

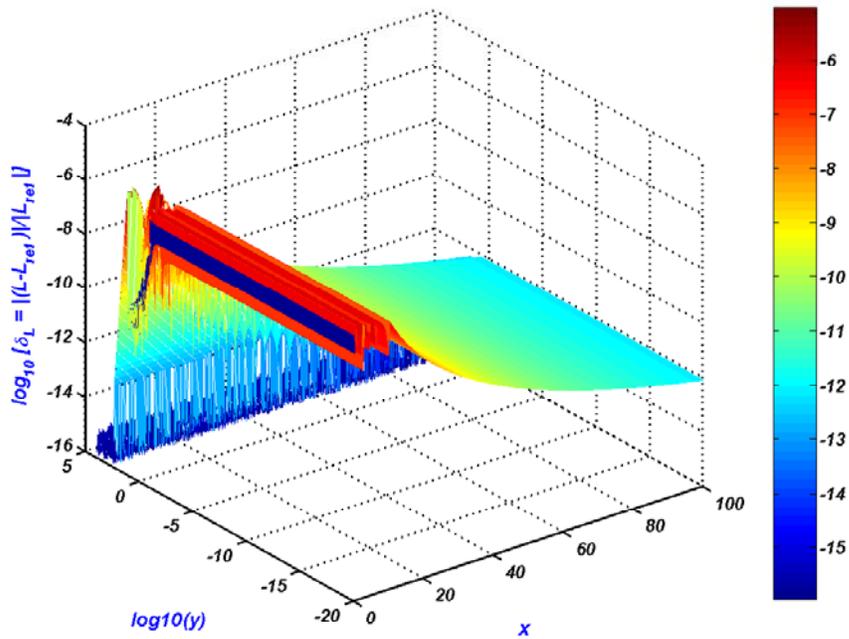

(b)

Figure 2: Absolute relative error resulting from using the present proposed routine for calculating; (a) the real part, and (b) the imaginary part of the Faddeyeva function using Algorithm 916 as a reference.

Figure 3 shows results from calculating $\partial V(x,y)/\partial x$ near the real axis ($y=10^{-20}$) from the present approximation as compared to Algorithm 916 and Humlíček's $w4(z)$ algorithm. It is clear from the figure that the present routine does not suffer from the loss of accuracy problem near the real axis shown by Humlíček's $w4(z)$ algorithm.

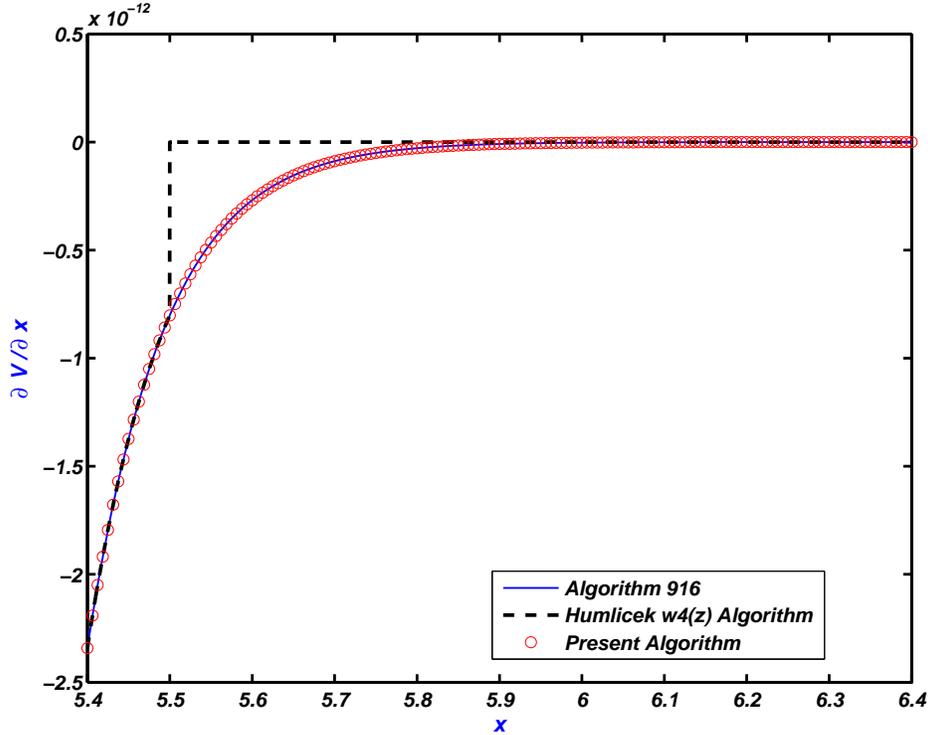

Figure 3: $\partial V(x,y)/\partial x$ as calculated by the present approximation, by *Algorithm 916* and by Humlíček's $w4(z)$ algorithm, for $y=10^{-20}$.

## 3. MATLAB AND FORTRAN IMPLEMENTATIONS AND PERFORMANCE RESULTS

The algorithm for our proposed method is described in Figure 1 and Table 1 and we have implemented this in both *Matlab* and *Fortran*. Performance tests and efficiency considerations are considered and discussed below.

### 3.1. Matlab Implementation

The proposed algorithm is implemented as a *Matlab* function *"wz(z)"* and has been tested for different regions in the computational domain. An efficiency comparison was performed between the present *Matlab* function and Humlíček's w4 algorithm using the same coding structure. The two functions were extensively tested and run over different regions of the computational domain. In addition to being marginally more accurate, and being safe from the loss of accuracy problem detected with Humlíček's w4 algorithm, the present routine was also found to be faster; see Table 2.

**Table 2:** Speed comparison between *Matlab* implementations of the present routine and Humlíček's w4 algorithm.

| Algorithm | Average time per evaluation (μs) | | | | Claimed accuracy and comments |
|---|---|---|---|---|---|
| | Case 1[*] | Case 2[*] | Case 3[*] | Case 4[*] | |
| *w4(z)* Humlíček [1982] | 0.1656 | 0.1715 | 0.2176 | 0.3481 | $<9.6\times10^{-5}$ (Loss of claimed accuracy for $y\leq10^{-6}$) |
| *wz(z)* Present Routine | 0.1290 | 0.1361 | 0.1712 | 0.2664 | $<4.0\times10^{-5}$ |

[*]Case 1: *y=logspace(-5,5,71)*, *x=linspace(-500,500,40001)*; Case 2: *y=logspace(-20,4,71)*, *x=linspace(-200,200,40001)*; Case 3: *y=logspace(-5,5,71)*, *x=linspace(-10,10,40001)*; Case 4: *y=logspace(-20, log₁₀(6),71)*, x randomly generated with $|z|^2\leq36$. In *Matlab* "*logspace(x1, x2, N)*" generates a vector of *N* points logarithmically equally spaced between $10^{x1}$ and $10^{x2}$, while "*linspace(x1, x2, N)*" generates a vector of *N* points linearly equally spaced between $x_1$ and $x_2$.

### 3.2. Fortran Implementation

The algorithm was also implemented as a *Fortran* elemental subroutine "*wz_rk(z,w)*" contained in a *Fortran* module "*wz_mod_rk*" that can be run using single or double precision arithmetic depending on the choice of an integer parameter "*rk*" in a subsidiary module "*set_rk*". The elemental subroutine operates on a single dummy complex argument, *z*, and returns the corresponding Faddeyeva function value as a complex argument, *w*. In addition, partial derivatives of the real part of the Faddeyeva function, "*dVdx*" and "*dVdy*", may be returned as optional outputs. The routine may be invoked with arrays as actual arguments where it will be applied element-wise, with a conforming array return value. Extra performance gains may be obtained by performing some of the computations using real arithmetic. For example, by replacing the term *exp(-z²)* found in the Humlíček's expression used herein for region *V* by *exp(-x²)* for the region $62.0>|z|^2\geq2.5$ & $y^2<10^{-13}$ while maintaining the claimed accuracy.

The execution time of the present *Fortran* routine is compared with Humlíček's w4 algorithm. The comparison has been performed for the same four datasets reported for the *Matlab* performance comparison above. The results of the comparison, using double precision (*Fortran-d*) and single precision (*Fortran-s*) are given in Table 3. As can be seen from the table, the present routine is consistently faster than the w4 algorithm when run using double precision arithmetic. For single precision computations, the present routine is also faster for all four datasets and is more than twice as fast for dataset 1.

Table 3: Speed comparison between *Fortran* implementations of the present routine and Humlíček's w4 algorithm. The values given are for the calculation of the function only (i.e., no derivative calculations) and have been generated using Intel Visual Fortran Compiler Professional for applications running on IA-32, Version 11.1.038

| Algorithm | | Average time per evaluation (μs) | | | |
|---|---|---|---|---|---|
| | | Case 1 | Case 2 | Case 3 | Case 4 |
| *w4(z)* Humlíček [1982] | *Fortran-d* | *0.0115* | *0.0128* | *0.0296* | *0.0707* |
| | *Fortran-s* | *0.0119* | *0.0128* | *0.0255* | *0.0560* |
| *wz(z)* Present Routine | *Fortran-d* | *0.0077* | *0.0108* | *0.0270* | *0.0599* |
| | *Fortran-s* | *0.0056* | *0.0077* | *0.0228* | *0.0509* |

## 4. CONCLUSIONS

A simple, efficient, and relatively accurate ($<4.0\times10^{-5}$) approximation for the computation of the Faddeyeva function is presented. The routine can be easily implemented in any computational framework. *Matlab* and *Fortran* implementations are provided which exhibit improved accuracy and better efficiency when compared to Humlíček's w4 algorithm, which is widely used for massive evaluation of the function in the literature.

**ACKNOWLEDGMENTS**

The author would like to thank the anonymous reviewers for valuable and constructive suggestions.
The insightful comments and suggestions received from Mr. Van Snyder (Jet Propulsion Laboratory, Pasadena, CA) and Dr. Tim Hopkins (TOMS Algorithm Editor) during early stages of the present manuscript are real contributions to this work and are particularly appreciated.